\documentclass[a4paper,fleqn,usenatbib]{mnras}

\usepackage{txfonts}

\usepackage[T1]{fontenc}
\usepackage{ae,aecompl}

\usepackage{graphicx}	
\usepackage{amstext}	
\usepackage{amssymb}	

\newcommand{\NHI}{$N\rm _{H\,{\sevensize I}}$}

\newcommand{\lya}{Ly$\alpha$}

\newcommand{\sbline}{$\rm erg~s^{-1}~cm^{-2}~arcsec^{-2}$}

\newcommand{\cW}{$C_{\rm W}$} 
\newcommand{\cE}{$C_{\rm E}$} 
\newcommand{\eagle}{{\sc eagle}}

\title[Galaxy assembly at $z\approx3.25$ in an extended structure]{Witnessing galaxy assembly in an extended $z\approx 3$ structure}

\author[Fumagalli et al.]{Michele Fumagalli$^{1,2}$\thanks{E-mail: michele.fumagalli@durham.ac.uk},
  Ruari Mackenzie$^{2}$, James Trayford$^{1}$, Tom Theuns$^{1}$, \and
  Sebastiano Cantalupo$^{3}$, Lise Christensen$^{4}$, Johan P.~U. Fynbo$^{4}$, Palle M{\o}ller$^{5}$, \and
  John O'Meara$^{6}$, J. Xavier Prochaska$^{7,8}$, Marc Rafelski$^{9}$, Tom Shanks$^{2}$
  \\
  $^{1}$Institute for Computational Cosmology, Durham University, South Road, Durham, DH1 3LE, UK \\
  $^{2}$Centre for Extragalactic Astronomy, Durham University, South Road, Durham, DH1 3LE, UK \\
  $^{3}$Institute for Astronomy, ETH Zurich, Wolfgang-Pauli-Strasse 27, 8093 Zurich, Switzerland\\
  $^{4}$Dark Cosmology Centre, Niels Bohr Institute, University of Copenhagen, Juliane Maries Vej 30, DK-2100 Copenhagen, Denmark\\
  $^{5}$European Southern Observatory, Karl-Schwarzschild-Strasse 2, 85748, Garching, Germany\\
  $^{6}$Department of Chemistry and Physics, Saint Michael's College, One Winooski Park, Colchester, VT 05439, USA \\
  $^{7}$Department of Astronomy and Astrophysics, University of California, 1156 High Street, 
  Santa Cruz, CA 95064 USA \\
  $^{8}$University of California Observatories, Lick Observatory 1156 High Street, Santa Cruz, 
  CA 95064 USA \\
  $^{9}$ Space Telescope Science Institute, Baltimore, MD, USA \\
}

\pubyear{\the\year}

\begin{document}
\label{firstpage}
\pagerange{\pageref{firstpage}--\pageref{lastpage}}
\maketitle
\begin{abstract}
  We present new observations acquired with the Multi Unit Spectroscopic Explorer instrument on the Very Large
  Telescope in a quasar field that hosts a high column-density  damped Ly$\alpha$ absorber (DLA) at $z\approx 3.25$.
  We detect Ly$\alpha$ emission from a nebula
  at the redshift of the DLA with line luminosity $(27 \pm 1)\times 10^{41}~\rm erg~s^{-1}$, which extends
  over $37\pm 1~\rm kpc$ above a surface brightness limit of $6\times 10^{-19}~$\sbline\ at a projected
  distance of $30.5 \pm 0.5 ~\rm kpc$ from the quasar sightline.
  Two clumps lie inside this nebula, both with Ly$\alpha$ rest-frame equivalent width $> 50~\rm \AA$ and with
  relative line-of-sight velocities aligned with two main  absorption components seen in the DLA spectrum. 
  In addition, we identify a compact galaxy at a projected distance of $19.1 \pm 0.5 ~\rm kpc$ from the
  quasar sightline. The galaxy spectrum is noisy but consistent with that of a star-forming galaxy at the DLA redshift.
  We argue that the Ly$\alpha$ nebula is ionized by radiation from star formation inside the two clumps,
  or by radiation from the compact galaxy. In either case, these data imply the presence of a structure with
  size $\gg 50$ kpc inside which galaxies are assembling,  a picture consistent with galaxy formation in groups and filaments as predicted by  cosmological simulations such as the \eagle\ simulations.
\end{abstract}




\section{Introduction}

In the current cosmological paradigm, galaxies form in overdense regions of the Universe when gas is
funnelled from the intergalactic medium inside dark matter halos, where it settles into disks and cools
to form stars \citep{white1978,blumenthal1984}. A distinctive feature of this model is the clustering
of galaxies on scales ranging from individual halos of tens of kiloparsecs,
up to structures on scales of few megaparsecs. This picture is supported by hydrodynamic simulations,
as well as by the statistics recovered from observations of the starlight emitted by young galaxies
that populated the Universe more than ten billion years ago.

Direct observations of the
link between hydrogen and star formation within and around galaxies provide key insight into the astrophysical
processes that regulate galaxy evolution in this hierarchy of
groups and filaments \citep[e.g.][]{warren1996,weatherley2005,cucciati2014,mf2016,bielby2016,peroux2017,cai2017}.
However, due to the difficulty in mapping neutral hydrogen in emission at high redshift,
until recently it has proven challenging to study in detail the connection between gas and galaxies inside overdensities
beyond $z\gtrsim 2$.

Most of our knowledge of the neutral gas content of the high-redshift Universe originates from the study of damped Ly$\alpha$
absorbers (DLAs), which are detected in absorption against bright background sources \citep{wolfe2005}.
Defined as systems with \ion{H}{I} column density $ N_{\rm HI} \ge 2\times 10^{20}~\rm cm^{-2}$, DLAs trace the bulk of the
neutral hydrogen in the distant Universe and are the major repository of the fuel for the formation of the stars seen at
present days \citep[e.g.][]{prochaska2009,noterdaeme2009}. Many pieces of evidence link DLAs to star formation in high-redshift
galaxies. For instance, the high column density of DLAs
provides the necessary conditions for hydrogen to shield from the ambient ionizing radiation, potentially allowing the condensation of a cold and molecular phase that is thought to be essential for the formation of new stars \citep{krumholz2009,neeleman2015,noterdaeme2015,rafelski2016}. Furthermore, DLAs contain significant amounts of heavy elements with a mean metallicity of $\log Z/Z_\odot \approx -1.5$ at $z \gtrsim 2$,
which is more than an order of magnitude above what is found in the intergalactic medium
\citep[e.g.][]{prochaska2003,rafelski2012,rafelski2014}.

However, establishing a connection between DLAs and star-forming galaxies has been a non-trivial task. Despite decades of searches, only tens of DLAs have been associated directly to counterparts in emission, particularly thanks to recent campaigns  \citep[e.g.][]{moller2004,fynbo2010,noterdaeme2012,peroux2012,fynbo2013,neeleman2017,krogager2017}. This low detection rate
across the entire DLA population is generally attributed to the difficulties of imaging faint emission
from stellar populations or hydrogen recombination lines at high redshifts. It also likely reflects the fact that
only a fraction of the DLA population is directly connected to active {\it in situ} star formation,
or that only a subset of DLAs reside near galaxies which are sufficiently bright for direct detection
\citep[such as high column-density and/or metal-rich DLAs;][]{fynbo2008,moller2013,christensen2014,mf2015,joshi2017,krogager2017}.

Together with recent advancements at infrared and millimetre wavelengths \citep[e.g.][]{peroux2012,neeleman2017}, 
the deployment of the Multi Unit Spectroscopic Explorer \citep[MUSE;][]{bacon2010} at the Very Large Telescope (VLT) represents a breakthrough to address the long-standing issue of connecting DLAs to galaxy counterparts.
Indeed, MUSE allows us to image Ly$\alpha$ emission within $\approx$250 kpc from the position of known DLAs to
flux limits of $\approx 10^{-18}~\rm erg~s^{-1}~cm^{-2}$ in exposure times of only few hours, thus enabling
deep searches of DLA host galaxies without the need of pre-selecting targets for spectroscopic
follow-up observations.

In this work, we present new MUSE observations of a field centred on the quasar J$025518.58+004847.6$ at $z_{\rm qso}\approx 3.996$, where high-resolution absorption spectroscopy reveals the presence of a high-column density
DLA at the redshift $z_{\rm dla} = 3.2552 \pm 0.0001$ \citep{prochaska2001,mf2014}.
This target has been selected from a larger parent sample of quasar fields with available deep optical imaging
(see Sect.~\ref{sec:data}). The layout of this paper is the following. Details on the data acquisition and reduction are presented in Sect.~\ref{sec:data}, while the analysis and modelling of these observations are in Sect.~\ref{sec:analysis}, Sect.~\ref{sec:model}, and Sect.~\ref{sec:cfreagle}. Summary and conclusions are presented in Sect.~\ref{sec:disc}. Throughout this work, we use the Planck 2015 cosmological parameters \citep[$\Omega_{\rm m}=0.307$, $H_0=67.7~\rm km~s^{-1}~Mpc^{-1}$;][]{planck2016} and we quote magnitudes in the AB system.

\begin{figure*}
  \centering
  \includegraphics[scale=0.45]{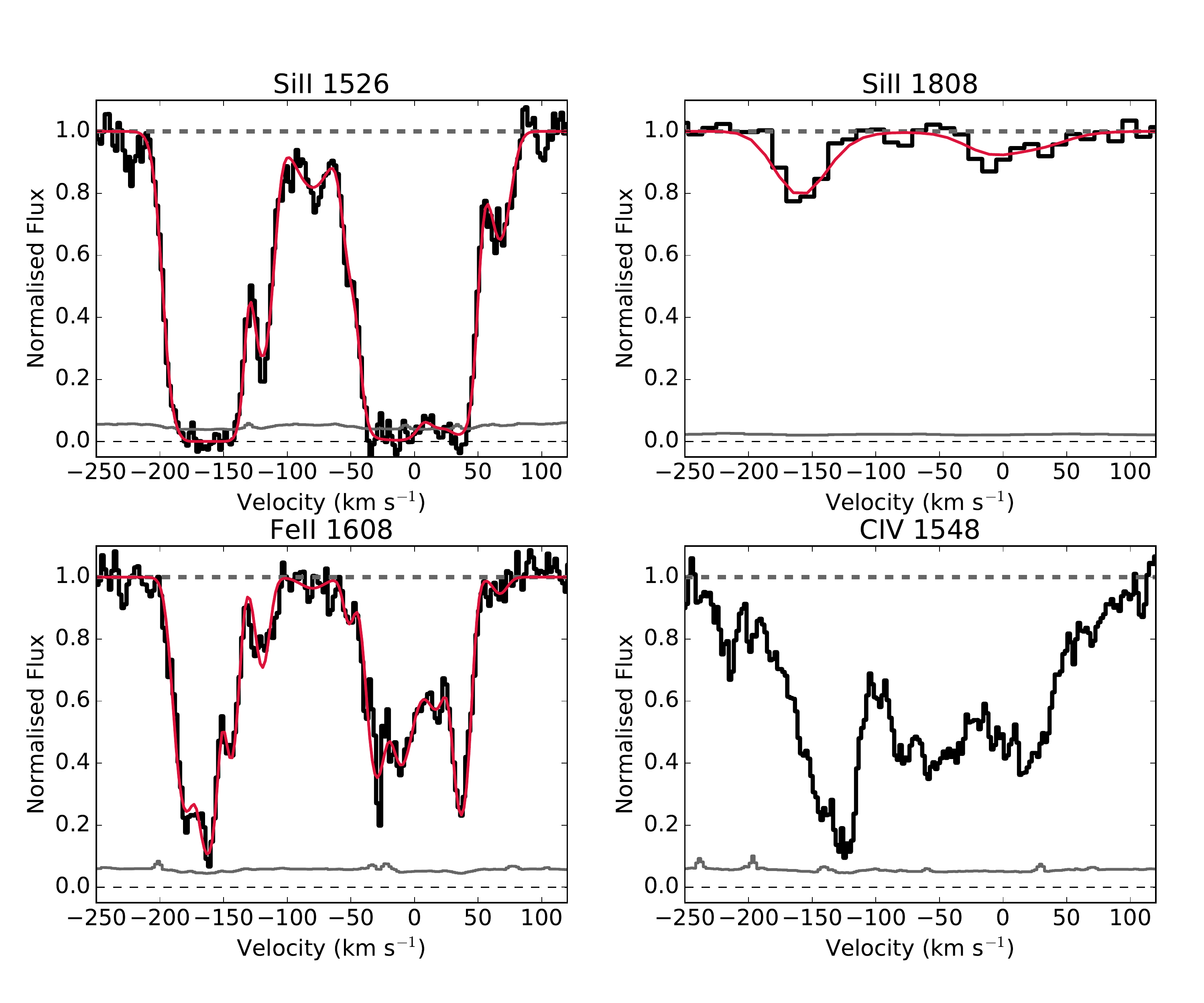}
  \caption{Gallery of selected metal transitions associated with the $z_{\rm dla} = 3.2552 \pm 0.0001$ DLA.
    Data from HIRES or X-SHOOTER (in the case of \ion{Si}{II}~$\lambda1808$) are shown in black,
    together with the associated $1\sigma$ error on the flux (grey solid line).
    The red solid line shows a multi-component Voigt profile fit to the low ionization lines.
    The grey dashed line at the top of each panel marks the
    continuum level. Two main components separated by $155 \pm 6 ~\rm km~s^{-1}$ are clearly visible, each with sub-components. 
  }\label{fig:dlalines}
\end{figure*}

\section{Data acquisition and reduction}\label{sec:data}
The quasar field J$025518.58+004847.6$ is one of the six fields selected
from the parent sample of quasars included in our imaging survey
designed to probe the {\it in situ} star formation rate (SFR) of high-redshift DLAs
\citep{omeara2006,mf2010,mf2014,mf2015}.
 The six fields have been selected to host DLAs at $z>3$, the redshift for which Ly$\alpha$ enters the wavelength range
covered by MUSE, with no other constraints on the physical properties of the absorbing gas (e.g. on metallicity).
MUSE observations for this sub-sample have, at the time of writing, been recently completed at VLT as part of the ESO
programmes 095.A-0051 and 096.A-0022 (PI Fumagalli). The line of sight presented in this work is the first field we
have analysed, and the entire sample will be presented in a forthcoming publication. 

MUSE observations of the quasar field J$025518.58+004847.6$ have been conducted during the nights 17-20 September
2015 in a series of 1500~s exposures totalling 2.5 hours on source, under good seeing (requested to be $\le 0.8''$) and clear
sky, but with non-photometric conditions.
Data have been reduced using the standard ESO MUSE pipeline  \citep[v1.6.2;][]{wei14} that 
includes basic data processing (bias subtraction and flat-fielding), plus wavelength and photometry calibrations.
Throughout this analysis, barycentric corrections have been applied to the MUSE data and the wavelengths have been
converted into vacuum for consistent comparisons with the spectroscopic data described below.
We further post-process the resulting datacubes to enhance the quality of sky subtraction and
flat-fielding using the {\sc CubExtractor} code ({\sc CubEx}, Cantalupo in prep.), following our earlier work
\citep{borisova2016,mf2016}.

The final data product is a cube that samples
the instrument field of view of $1 \times 1$ arcmin$^2$ in pixels of size $0.2$ arcsec
and, for each pixel, contains a spectrum covering the wavelength range $4750-9350~$\AA\ in bins of $1.25~$\AA. 
Following comparisons with photometric data from the Sloan Digital Sky Survey \citep{sdssIII}, we apply a small
correction factor of $1.12$ to the flux calibration to account for low levels of atmospheric extinction during the
observations.
We further correct fluxes to remove Milky Way dust extinction in the direction of our observations following
\citet{schlafly2011}.
At $\lambda \approx 5170~$\AA, which corresponds to the Ly$\alpha$ wavelength at the redshift of the DLA, we achieve an effective image quality of $\approx 0.6$ arcsec, and a noise level of
$\approx 6\times 10^{-19}~\rm erg~s^{-1}~cm^{-2}~\AA^{-1}~arcsec^{-2}$ (root mean square).

Throughout this analysis, we further make use of broad-band imaging data collected with the {\it Hubble Space Telescope} (HST) in the  F390W filter using Wide Field Camera 3, and
in the $u'$, $V$, and $R$ filters using the LRIS camera at the Keck Telescope.
The observations and reduction techniques for these data are described in \citet{mf2014}.
Finally, we use archival spectroscopic data on the DLA from the HIRES instrument at the Keck Telescope
and the X-SHOOTER instrument at VLT. These data are described in \citet{prochaska2001} and \citet{lopez2016}. 
The HIRES spectrum covers the wavelength range $5800-8155~$\AA\ with a signal-to-noise ratio ($S/N$) of $\approx 15$
per pixel and a spectral resolution of $\approx 6~\rm km~s^{-1}$. The X-SHOOTER spectrum covers instead the
wavelength range $3100-18000~$\AA\ with a $S/N\approx 30$
per pixel at a resolution of $\approx 40-60~\rm km~s^{-1}$, depending on the instrument arm.

\begin{figure*}
  \centering
  \includegraphics[scale=0.45]{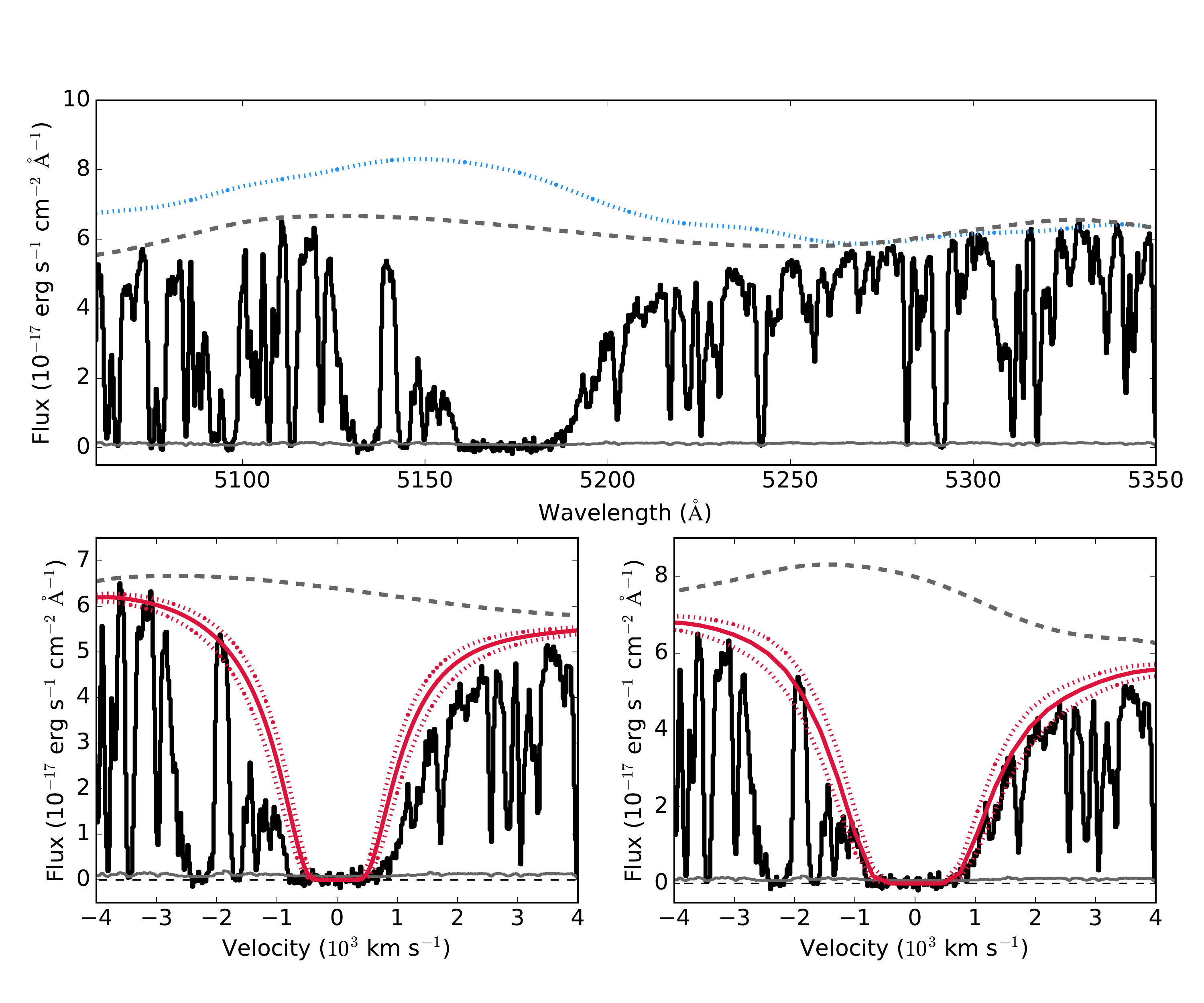}
  \caption{({\it Top panel}) X-SHOOTER spectrum of the J$025518.58+004847.6$ quasar (black),
    together with the associated error on the flux (grey solid line). The \ion{H}{I} profile of the
    $z_{\rm dla} = 3.2552 \pm 0.0001$ DLA is visible at $\lambda \approx 5173~\rm \AA$.
    The grey dashed line marks the local continuum, computed with a spline function that
    interpolates regions deemed free from absorption in the Ly$\alpha$ forest.
    The blue dotted line shows instead the assumed continuum from the
    quasar template by \citet{telfer2002}, which is normalised to reproduce
    regions of the spectrum with no absorption.
    This continuum level deviates from the local one around $\approx 5150~\rm \AA$
    because of the presence of Ly$\beta$ and \ion{O}{VI}
    emission lines of the quasar.
    {\it (Bottom panels)}
    Closer view of the \ion{H}{I} profile of the DLA, together with a Voigt profile model
    (red solid line) and associated errors (red dotted lines).
    The two panels differ in the choice of the adopted continuum level (grey dashed lines),
    which is either based on a local determination (left panel) or on a quasar template
    (right panel) as described above. 
  }\label{fig:dlahi}
\end{figure*}

\section{Analysis of the imaging and spectroscopic data}\label{sec:analysis}

\subsection{The physical properties of the $z\approx 3.25$ DLA}\label{sect:DLA}

A gallery of selected absorption lines at the redshift of the $z\approx 3.25$ DLA,
which are chosen to be representative of low ionization transitions, is shown in
Figure~\ref{fig:dlalines}. The profiles of low ionization metal lines reveal the presence of two distinct
main components, each showing additional sub-components.
Velocity structure is also seen in \ion{C}{IV}, although this ion does not
track the low ionization transitions in velocity space, suggesting the presence of multi-phase
gas in this DLA \citep[e.g.][]{fox2007}.
By fitting Voigt profiles to the \ion{Si}{II}~$\lambda1526$ and the
\ion{Fe}{II}~$\lambda1608$ lines using the VPFIT code \citep{carswell2014}, we find a column-density weighted
redshift of $z_{1} = 3.2530 \pm 0.0001$ and $z_{2} = 3.2552 \pm 0.0001$ for the two main components.

Further, a Voigt profile centred at $z_2$ provides a good description of the Ly$\alpha$ absorption
line and, in the following, we assume $z_{2}$ as the fiducial redshift for the DLA. 
As shown in Figure~\ref{fig:dlahi}, the measurement of the \ion{H}{I} column density is complicated
by the fact that the Ly$\alpha$ transition at the DLA redshift lies next to the
\ion{O}{VI} and Ly$\beta$ emission lines of the quasar, a coincidence that makes our estimate of the
continuum level rather uncertain. For this reason, we derive at first a lower limit on the DLA column density
by assuming a ``local'' continuum, which we obtain with a spline function 
constrained to follow regions deemed free from absorption inside the Ly$\alpha$ forest
(grey dashed line in the top panel of Figure~\ref{fig:dlahi}). The resulting model for the DLA
(bottom-left panel of Figure~\ref{fig:dlahi}) yields a column density of
$\log(N_{\rm HI}/\rm cm^{-2}) > 20.55 \pm 0.10$, which is primarily constrained by the blue wing.

However, it is evident that additional absorption is present in the red side of the \ion{H}{I} transition.
Therefore, we construct a second model adopting the continuum level from the quasar template by
\citet{telfer2002}, which we normalise to reproduce the observed spectrum in regions not absorbed by the
Ly$\alpha$ forest (blue dotted line in the top panel of Figure~\ref{fig:dlahi}).
With this continuum, we derive a second model for the DLA 
with $\log(N_{\rm HI}/\rm cm^{-2}) = 20.85 \pm 0.10$ (bottom-right of Figure~\ref{fig:dlahi}),
which is primarily constrained by the  red wing of the absorption profile.
As the red wing is less affected by the quasar emission lines, our model is not very sensitive to the exact shape of the template. For instance, the template by \citet{vandenberk2001} yields a consistent model within errors. 
Qualitatively, this model appears to reproduce well the shape of the Ly$\alpha$ transition and hereafter, we assume $\log(N_{\rm HI}/\rm cm^{-2}) = 20.85 \pm 0.10$ as our best estimate for the
\ion{H}{I} column density. The quoted error does not reflect the systematic uncertainty on the continuum placement.

Before continuing with our analysis, we note that the low velocity separation of the two main metal components allows for a further decomposition of the \ion{H}{I} profile, with components at $z_1$ and $z_2$. Both in the cases of the 
local continuum determination and the continuum based on a quasar template, solutions with two components yield 
a total \NHI\ in agreement with the single-component analysis.
However, the lack of high-order hydrogen lines, which fall beyond the Lyman limit of a higher redshift absorber, makes an unambiguous  decomposition difficult. Further, sub-structures are evident within the metal lines beyond the two main components, suggesting that the physical interpretation of a two-cloud model is still limited. For this reason, we proceed with a single-component fit for \ion{H}{I}, which is nevertheless effective in describing the total amount of neutral hydrogen 
within this system.

\begin{figure*}
  \centering
  \includegraphics[scale=0.5]{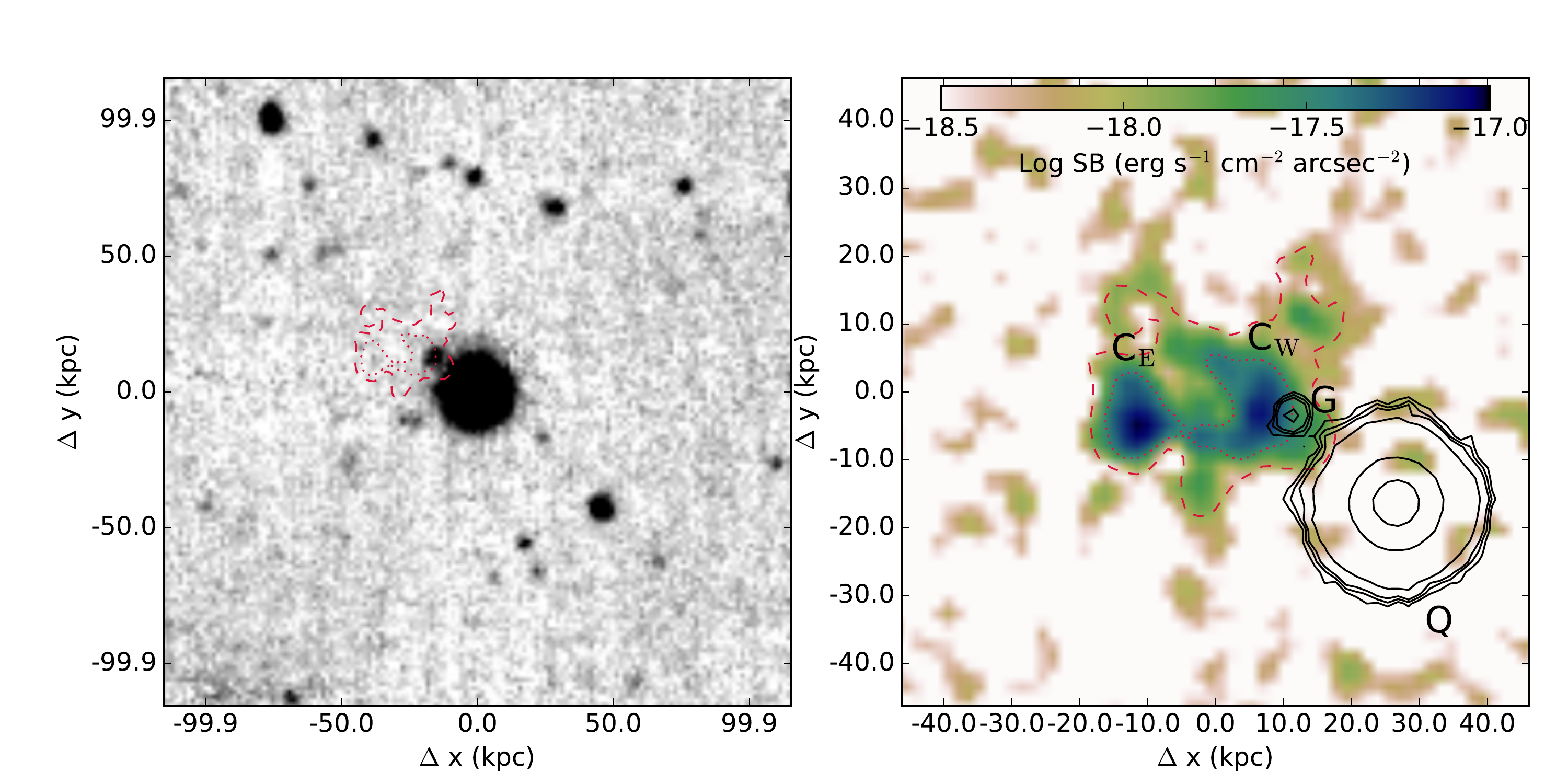} 
  \caption{({\it Left panel}) The white-light image reconstructed from the MUSE datacube
    showing the quasar at the centre, where absorption spectroscopy reveals the
    presence of a DLA with column density $\log N_{\rm HI} = 20.85 \pm 0.10 ~\rm cm^{-2}$ at
    $z_{\rm dla} = 3.2552 \pm 0.0001$. The red contours mark the $1\sigma$ and $5\sigma$
    surface brightness levels of the source detected in Ly$\alpha$ at the DLA redshift.
    North is upward and East to the left. 
    ({\it Right panel}) The pseudo narrow-band image of the Ly$\alpha$ emission at the DLA
    redshift, smoothed by 2 pixels. The red contours identify the $1\sigma$ and $5\sigma$
    surface brightness levels, respectively. MUSE observations reveal the presence of a $37 \pm 1~\rm kpc$
    structure, with two distinct clumps, labelled ``\cE'' and ``\cW''.
    The black contours mark the position of the quasar (labelled ``Q'') and the continuum-detected galaxy
    assumed at the DLA redshift (labelled ``G'').}\label{fig:dlaimg}
\end{figure*}

Strong metal lines in the DLA spectrum reveal that the absorbing gas is significantly enriched,
with the majority of the low-ionization transitions being saturated, with the exception of the weak
\ion{Si}{II}~$\lambda1808$ transition.
After correcting the spectrum for partial blending arising from the telluric band between $\approx 7600-7700~$\AA,
we apply the kinematic model derived from the \ion{Si}{II}~$\lambda1526$ line to the
\ion{Si}{II}~$\lambda1808$ transition, finding a column density of $\log(N_{\rm SiII}/\rm cm^{-2}) = 15.29 \pm 0.05$
which is consistent with earlier measurements based on the apparent optical depth method
\citep{prochaska2001,mf2014}.
Combined with the measurement of the \ion{H}{I} column density and assuming the relative abundance pattern of the Sun
from \citet{asplund2009}, we derive a metallicity of $\log Z/Z_\odot < -0.8 \pm 0.1$ if
$\log(N_{\rm HI}/\rm cm^{-2}) > 20.55 \pm 0.10$, or
$\log Z/Z_\odot = -1.1 \pm 0.1$ assuming the fiducial value of column density ($\log(N_{\rm HI}/\rm cm^{-2}) = 20.85 \pm 0.10$).
No ionization corrections are included, as they are known to be negligible for DLAs.
Similarly, no corrections for dust depletion have been applied, although we note
they may be non-negligible given the metallicity of this system \citep[up to $\approx 0.5~\rm dex$, but
generally $\lesssim 0.2~\rm dex$ for Si;][]{rafelski2012,decia2016}.

From our analysis, we conclude that this system is enriched above the median value of $\log Z/Z_\odot \approx -1.5$
for the DLA population at $z\approx 2-3$ \citep{rafelski2012}.
Moreover, the line profiles exhibit a complex kinematic structure with two main velocity components
stretching over $>200~\rm km~s^{-1}$, placing this DLA in the upper end of the metallicity and kinematics
distribution of the DLA population at these redshifts \citep{ledoux2006,neeleman2013}.

\begin{figure}
  \centering
  \includegraphics[scale=0.5]{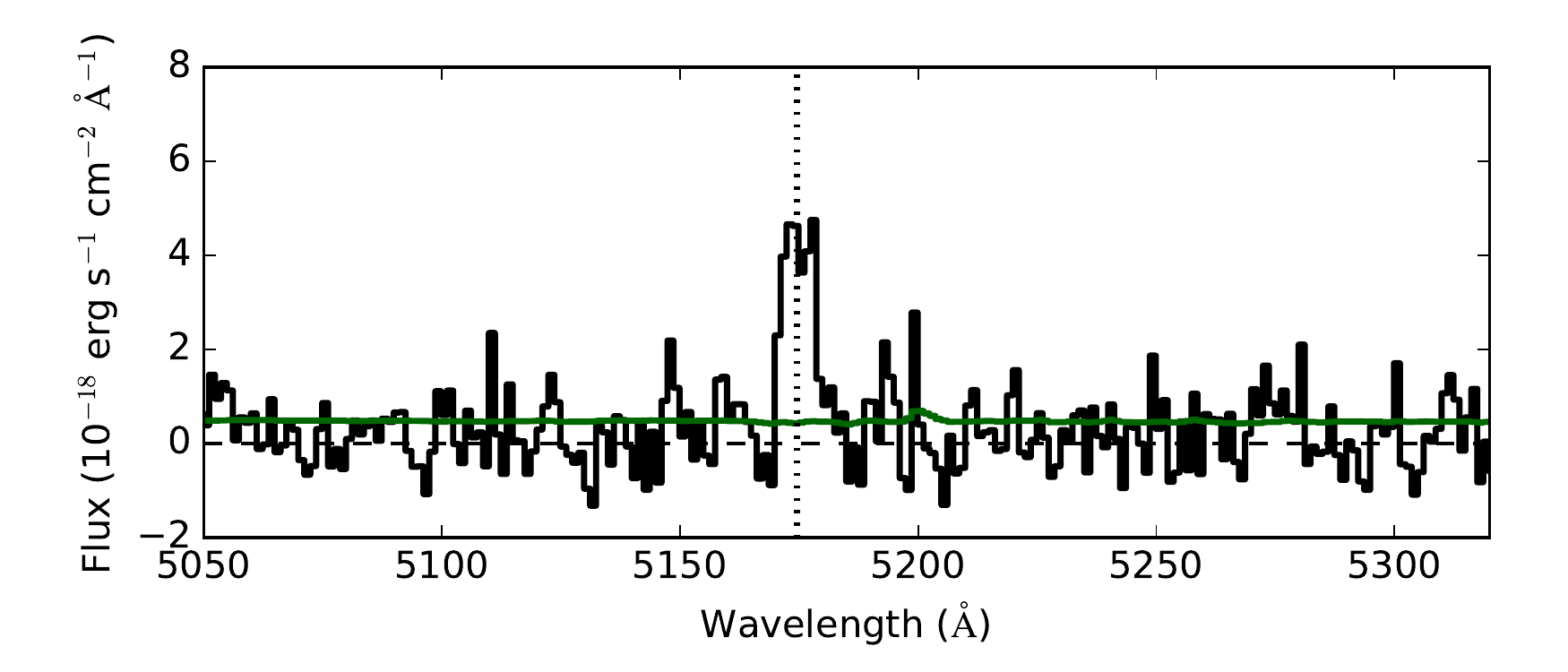} 
  \caption{Spectrum of the Ly$\alpha$ nebula, extracted from the MUSE cube in a mask that encompasses the full extent of the emission. The corresponding error array is shown in green. The vertical dotted line marks the mean redshift of the nebula.}\label{fig:neb1d}
\end{figure}

\begin{figure*}
  \centering
  \includegraphics[scale=0.8]{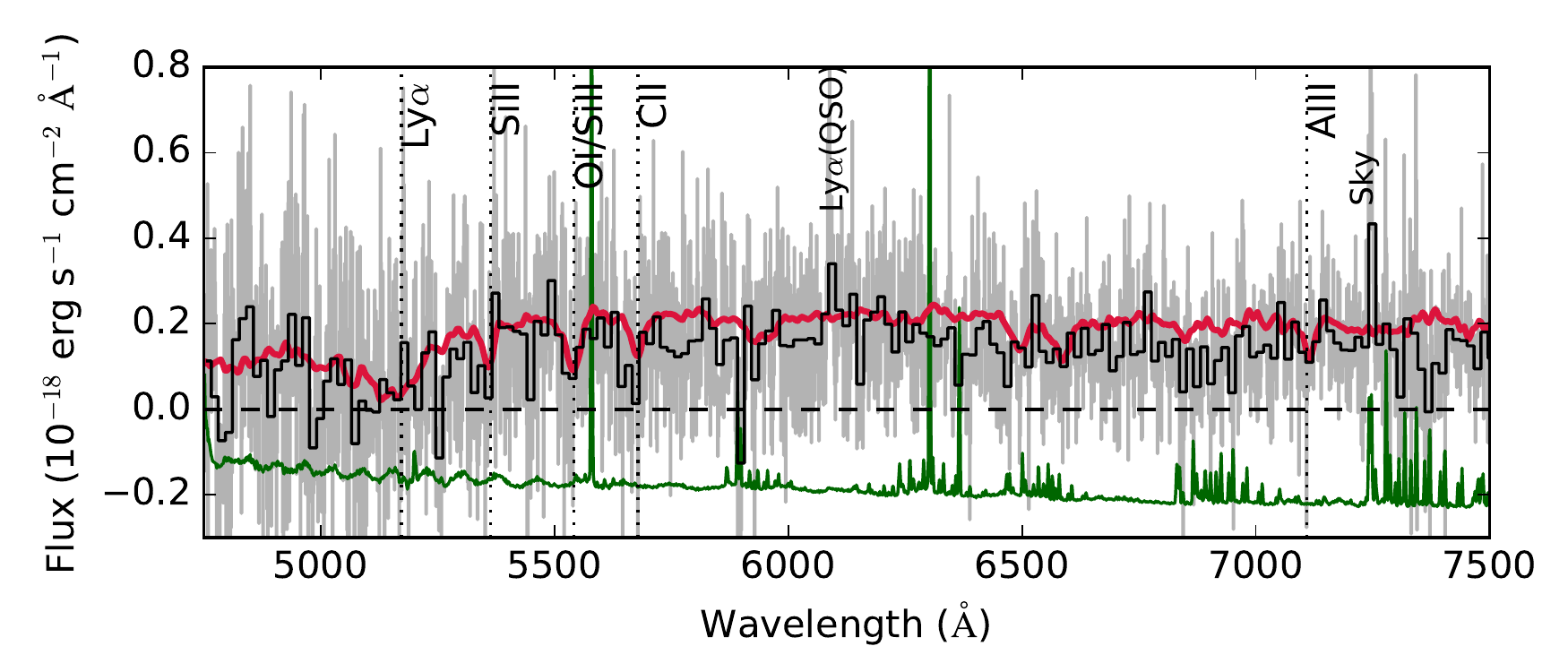} 
  \caption{Spectrum of galaxy G from MUSE data, before (grey) and after (black) rebinning by 12 spectral pixels
    (corresponding to 15\AA). The error array (green) has been offset for visualisation purposes.  The red line shows the composite LBG spectrum from \citet{bielby2013}. 
    Ly$\alpha$ and metal lines (as labelled) are tentatively detected in absorption.
    These spectral features, together with colour information and the source position, place this galaxy at a
    likely redshift $z_{\rm G} = z_{\rm dla}$.}\label{fig:sourceG}
\end{figure*}

\subsection{Search for galaxy counterparts}
The search for Ly$\alpha$ emitting galaxies inside the MUSE cube
is performed using the three-dimensional (3D) source extraction code
{\sc CubEx} (Cantalupo in prep.), following the techniques we developed for the analysis of
MUSE observations in quasar fields at $z\approx 3-4$ \citep[e.g.][]{mf2016}.

Briefly, after convolving the cube with a boxcar filter of 2 pixels on a side in the spatial
direction, we search for candidate sources which are detected with $S/N > 3$ within a minimum
volume of 30 voxels. After producing 3D segmentation maps of all the detected objects, we create pseudo narrow-band
images by projecting the cube along the wavelength direction as described in \citet{borisova2016}.
Next, we visually inspect these images and the associated 1D spectra to exclude both spurious sources
(cosmic rays and noise artefacts at the edge of the field of view) and lower-redshift line emitters,
which we identify by virtue of resolved [OII] doublets or multiple emission lines. 

In the end, our search for Ly$\alpha$ emitting sources 
within $\pm 500~\rm km~s^{-1}$ around the DLA redshift 
reveals the presence of an extended structure ($1\sigma$ level of $6\times 10^{-19}~$\sbline) near the quasar, with two
distinct clumps (hereafter the East ``$C_{\rm E}$'' and West ``$C_{\rm W}$'' clumps), as shown in Figure \ref{fig:dlaimg}.
A 1D spectrum extracted from a mask that encompasses the extent of the nebula (see Sect. \ref{sect:galaxies}) is in Figure~\ref{fig:neb1d}. The Ly$\alpha$ emission line is detected at high significance.

No other Ly$\alpha$ sources are detected to a flux limit of $> 2\times 10^{-18}~\rm erg~s^{-1}~cm^{-2}$
($3\sigma$ confidence level). Based on the completeness tests we performed in \citet{mf2016},
our search is assumed to be $\gtrsim 80\%$ complete to a flux of $\gtrsim 4.5\times 10^{-18}~\rm erg~s^{-1}~cm^{-2}$,
after accounting for the shorter exposure time of these observations.

The detected source is clearly associated with the DLA, being separated by only
$90 \pm 20 \rm ~km~s^{-1}$ in velocity space and having a centroid offset of $30.5\pm 0.5$ kpc in projection
with respect to the quasar sightline.
The velocity offset increases to $\approx 250~\rm km~s^{-1}$ if we assume $z_1$ instead of $z_2$
as the redshift of the DLA. In any case, we note that redshifts
of $\approx 300 \rm ~km~s^{-1}$ relative to systemic velocity are quite common for
the resonant Ly$\alpha$ transition in Lyman break galaxies \citep[LBGs; e.g.][]{steidel2010,bielby2011,rakic2011},
and are also seen in DLAs with detected Ly$\alpha$ emission \citep[e.g.][and reference therein]{srianand2016,joshi2017}.

Finally, we measure spectroscopic redshifts for the continuum-detected sources, following the procedures
detailed in \citet{mf2016}, which are here only briefly summarised. 
After generating a deep white-light image by collapsing the cube along the wavelength axis, we
run the {\sc SExtractor} code \citep{bertin1996}
with a detection area of 8 pixels and a threshold of $1.5\sigma_{\rm rms}$,
where $\sigma_{\rm rms}$ is the background root mean square. For each detected source, we then extract
the 1D spectrum, which is visually inspected by two of the authors (MF and RM) to assign a redshift.
Of all the galaxies to which we could assign a spectroscopic redshift
following this procedure (typically with $m_{\rm R} \lesssim 25$ mag), none is found to lie within
$\pm 500~\rm km~s^{-1}$ of the DLA.

Due to its close separation to the quasar in projection, source G (Figure \ref{fig:dlaimg})
requires an additional local background subtraction to correct for the contamination from the
quasar light in the wings of the point spread function. To account for this, we extract an annulus centred
on the quasar that encompasses the full width of galaxy G and contains $\approx 200$ pixels, without including the emission from the galaxy itself.
We next generate a composite spectrum of the pixels contained in this annulus, which we subtract from the source
spectrum.

As shown in Figure \ref{fig:sourceG}, the residual spectrum is very noisy. However, 
this source has a continuum emission consistent with a galaxy at $z\approx 3.2$ when
compared to an LBG template \citep{bielby2013}. The high-redshift nature of this galaxy is also confirmed by the 
lack of detection in the LRIS $u'$ filter ($m_{\rm u'} < 27.8~\rm mag$) and the marginal detection
in the HST F390W filter ($m_{\rm f390w} = 27.98 \pm 0.23~\rm mag$). Combined with the $R$ and $V$ magnitudes measured in the
Keck imaging ($m_{\rm V} = 26.19 \pm 0.24~\rm mag$; $m_{\rm R} = 25.33 \pm 0.18~\rm mag$),
we derive colours of $U-V = 1.8\pm 0.3~\rm mag$ and $V-R = 0.9\pm 0.3~\rm mag$, which are consistent with those of
$z\approx 3$ LBGs \citep{steidel2003,bielby2011}.
Moreover, after rebinning the spectrum, \ion{H}{I} in absorption and perhaps metal lines at the DLA redshift are tentatively detected. 
We observe no Ly$\alpha$ emission from this galaxy, but this is not unusual even for star-forming LBGs \citep{steidel2011}.
The best-fit redshift obtained comparing the spectro-photometry of galaxy G with the spectral energy distribution of an LBG is $z_{\rm G} = 3.22\pm0.03$.

Given the presence of neutral hydrogen at the DLA position
and that neutral hydrogen is likely present also at the location of the Ly$\alpha$ emission,
it is possible that the features seen in the spectrum of galaxy G originate
from foreground absorption against the UV continuum of a source at $z > z_{\rm dla}$ \citep[see][]{cooke2015}.
For this reason, and given the marginal quality of the data, we cannot determine a secure
redshift for galaxy G. However, the colour information combined with the source position near the nebula
makes galaxy G a compelling candidate for a system within the same structure that gives rise
to the DLA and the extended Ly$\alpha$ emission. Thus, in the following, we consider this galaxy
at the redshift $z_{\rm G} = z_{\rm dla}$. Future follow-up observations targeting
rest-frame optical lines in the IR will test our working hypothesis. 

\begin{figure}
  \centering
  \includegraphics[scale=0.53]{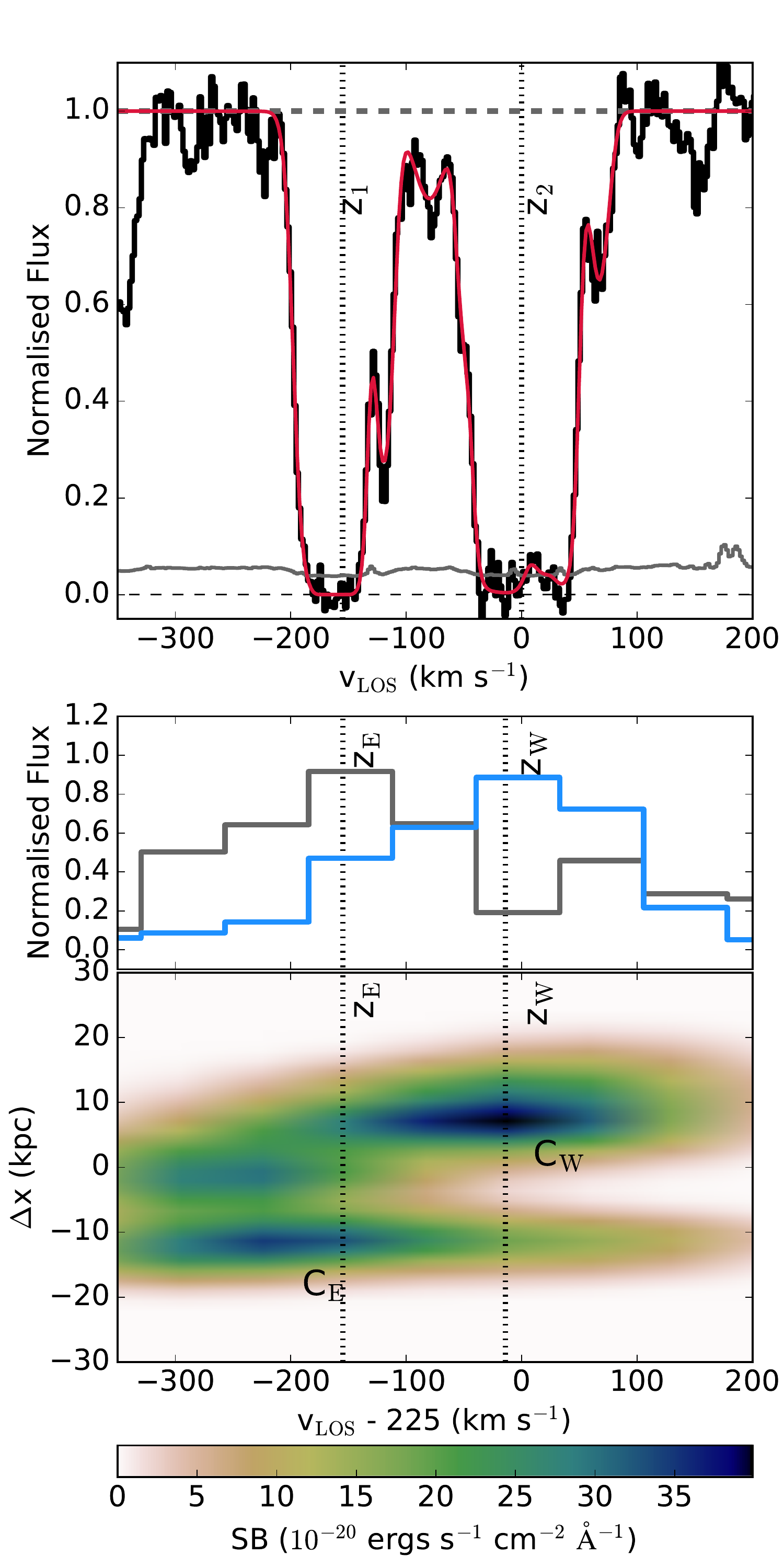} 
  \caption{{\it (Top panel)} The \ion{Si}{II} line profile from HIRES data as in Figure~\ref{fig:dlalines},
    with the redshifts of the two main components marked by vertical dotted lines. {\it (Bottom panel)}
    Ly$\alpha$ narrow-band velocity/position map reconstructed from the MUSE datacube, with the inset
    showing the projected 1D spectrum. The grey and blue lines in the inset refer to the \cE\ and \cW\ clumps, respectively. The redshifts of the two clumps are also marked with vertical dotted lines.
    Once an offset of $\Delta v = -225~\rm km~s^{-1}$ is applied to align the East clump
    with the first absorption component, the absorption and emission spectra appear to track each other
    in velocity space.}\label{fig:lyakin}
\end{figure}

\subsection{Properties of the detected galaxies}\label{sect:galaxies}

With a flux of $f_{r} = (16 \pm 1)\times 10^{-20}~\rm erg~s^{-1}~cm^{-2}~\AA^{-1}$ measured in
an image generated from the MUSE cube convolved with the SDSS $r$-band filter, galaxy G is forming stars at a rate
of $3.7 \pm 0.2~\rm M_\odot~yr^{-1}$, when adopting the calibration described in \citet{mf2010}
and neglecting internal dust extinction.  The impact parameter between the quasar sightline
and galaxy G is $19.1 \pm 0.5~\rm kpc$, sufficiently large to make it quite unlikely that
we are probing directly the galaxy interstellar medium in absorption.

As our data cover only the rest-frame UV ($\lesssim 2000~\rm \AA$), we do not have a direct estimate of the
galaxy stellar mass. Using empirical correlations between the SFRs and stellar masses
of LBGs between $z\approx 2-4$, galaxy G is expected to have a stellar mass of $\approx 10^9~\rm M_\odot$
\citep{whitaker2014,salmon2015}. Conversely, the stellar mass expected from scaling relations
calibrated on known DLAs is a factor $\approx 5$ higher, as it is predicted to be
$\approx 5 \times 10^{9}~\rm M_\odot$ for a DLA with metallicity of
$\log Z/Z_\odot \approx -1.1$ at the observed impact parameter \citep{moller2013,christensen2014}.
However, the large scatter in both relations combined with the unknown dust extinction for galaxy G makes reliable
comparisons between these methods difficult without additional observations that target the
rest-frame optical and IR wavelengths. It is also interesting to note that this detection occurs in a metal-rich
and high column-density DLA, in agreement with the trend that has emerged from searches of DLA counterparts completed
in the past decades \citep{krogager2017}.

Focusing on the extended Ly$\alpha$ emission next,  we use the surface brightness map
extracted from the MUSE cube to define a region that encloses the full source
(with a contour at the $1\sigma$ level of $6\times 10^{-19}~$\sbline), and the two clumps (with a contour at the $5\sigma$ level).
These apertures are shown in Figure~\ref{fig:dlaimg} and are used throughout this analysis to compute Ly$\alpha$
line fluxes and for aperture-matched broadband photometry.
Under these assumptions, the two knots \cE\ and \cW\ are separated
by $16.5\pm 0.5~\rm kpc$, and they contain respectively $(28\pm 2)\%$ and $(35\pm 3)\%$ of the total line flux of
$(2.8\pm 0.1)\times 10^{-17}~\rm erg~s^{-1}~cm^{-2}$. As shown in Figure~\ref{fig:dlaimg}, these clumps are embedded
in an extended component that stretches for $37 \pm 1$ kpc on a side. To date, this is one of the largest
Ly$\alpha$ emitting structures associated with a DLA \citep[see e.g.][]{kashikawa2014}.

When examining the line-of-sight velocity derived from the MUSE cube,
differential velocity is seen across the structure, with the two clumps being separated
by $\Delta v_{\rm E-W} = 140\pm 20~\rm km~s^{-1}$ in velocity space (Figure~\ref{fig:lyakin}).
This velocity separation is consistent, within errors, with the velocity separation inferred
from the two main absorption components ($\Delta v_{1-2} = 155 \pm 6~\rm km~s^{-1}$), suggesting that the
Ly$\alpha$ emission from the clumps follows the absorption profile of the DLA in velocity space.
We note, however, that radiative transfer effects modulate the Ly$\alpha$ line profile,
potentially altering the relative velocity separation that is intrinsic for the two clumps. 

The kinematic alignment is further highlighted in Figure~\ref{fig:lyakin}, where we compare the emission and
absorption spectra. After shifting the Ly$\alpha$ emission by $\Delta v = -225 ~\rm km~s^{-1}$
to align  the first main absorption component with the redshift of $C_{\rm E}$,
the DLA profile and the Ly$\alpha$ line emission show remarkable similarity, reinforcing the
hypothesis of a physical association between the two clumps and the DLA. 
Physically, $\Delta v$ reflects a combination of
the offset between emission and absorption arising from radiative transfer effects and
intrinsic velocity difference due to peculiar motions along the line of sight.
With current observations, we cannot separate these two effects, but future IR observations that target rest-frame
optical lines will be able to further constrain the kinematics of this system.

Focusing on the extended Ly$\alpha$ emission, we note that it is largely non-overlapping with galaxy G.
When masking this compact source, our analysis of deep imaging data from Keck, VLT, and HST
reveals that no continuum emission is detected from the rest-frame wavelength interval $821-2198$~\AA, to a limit of
$<1.7\times 10^{-19}~\rm erg~s^{-1}~cm^{-2}~$\AA$^{-1}$\ ($2\sigma$) that is calculated on the deepest imaging obtained
by collapsing the MUSE cube along the wavelength axis. Hence, without considering galaxy G, the lack of extended UV
continuum across this structure sets a limit for the average SFR  of $< 5.1 ~\rm M_\odot~yr^{-1}$
for the entire system, and $<0.9 ~\rm M_\odot~yr^{-1}$ and $<1.3 ~\rm M_\odot~yr^{-1}$ for the E and W clumps.
These estimates do not account for possible dust-obscured components. 
The corresponding Ly$\alpha$ rest-frame equivalent width for the two clumps (E and W) are $>60.7$~\AA\ and  $>52.3$~\AA.
Here, non-detections are computed at $2\sigma$ confidence level accounting for correlated noise by measuring the
standard deviation of fluxes recovered in empty regions of the images within apertures of equal size to those defined above. With its UV flux, however, galaxy G contributes to the total equivalent width of the structure, which becomes
$41\pm 3~$\AA\ in the rest frame.

\section{Nature of the extended emission}\label{sec:model}

Our MUSE observations reveal an emitting structure of $\approx 40~\rm kpc$ on a side,
the centre of which lies at an impact parameter of $\approx 30~\rm kpc$ from the quasar sightline hosting
a high-column density DLA.
Further, a compact UV source is found at $\approx 20~\rm kpc$ in projection from the quasar sightline, near the nebula.
Combined, these pieces of evidence suggest the presence of a structure that contains neutral
hydrogen and stretches for over $\approx 50~\rm kpc$ in projection, giving rise to the DLA at the quasar position and
the nebula. This structure further hosts one or multiple galaxies (see below). 

A structure with these properties is reminiscent of the filamentary structures predicted by cosmological
simulations, at the intersection of which galaxies form. Indeed, recent hydrodynamic simulations
\citep{mf2011,vandevoort2012,rahmati2014} consistently predict the presence of extended ($\gg 100~\rm kpc$)
filaments which give rise to absorption systems with $\log (N_{\rm HI}/\rm cm^{-2}) > 17$ in the surroundings
of dark matter halos, and which host multiple satellite galaxies clustered to the main parent halo.
Thus, interpreted in this context, our observations are probing the denser gas distribution
either inside the circumgalactic medium (CGM) of galaxy G (with the caveat on the redshift discussed above) or inside a group of galaxies, with
the observed Ly$\alpha$ emission being powered by star formation.
These two scenarios are discussed in the reminder of this section.
 A more direct comparison between the observed system and the results of a recent cosmological hydrodynamic simulation follows in 
Sect.\ref{sec:cfreagle}.

\subsection{Photoionization from galaxy G}\label{sec:photo}

With an average surface brightness $\Sigma_{\rm Lya} = (2.20 \pm 0.08) \times 10^{-18}~\rm erg~s^{-1}~cm^{-2}~arcsec^{-2}$
within $\approx 30~\rm kpc$ from the star-forming galaxy G, the observed Ly$\alpha$ emission
may arise from photoionization and/or collisional ionization in the CGM
of this galaxy. Indeed, extended Ly$\alpha$ emission in excess of $\approx 10^{-18}~\rm erg~s^{-1}~cm^{-2}~arcsec^{-2}$
is typically seen around LBGs or low-mass star-forming galaxies and it is ubiquitously
predicted by numerical simulations \citep[e.g.][]{steidel2011,wisotzki2016,furlanetto2005,faucher2010,rosdahl2012}.

Considering the morphology and location of the Ly$\alpha$ emission, we note that a model in which galaxy G is
powering this nebula needs to account for both the off-centre position of the UV source
compared to the Ly$\alpha$ emitting structure and for the presence of substructures inside the nebula.
Both requirements can be satisfied, at least qualitatively. First,
offsets between the sites of star formation and regions of Ly$\alpha$ emission have been reported in the
literature \citep[e.g.][]{sobral2015}. Moreover, density fluctuations (or the presence of sub-halos)
in the CGM could account for the observed clumps.

The next condition that must be satisfied is whether, from an energetic point of view, galaxy G can power the observed
emission. Given the observed SFR, we estimate that this source produces ionizing radiation at a rate of
$\log (Q_{\rm H0}/s^{-1}) \approx 53.9$, according to a {\sc starburst99} model with constant star formation rate,
Geneva stellar tracks without rotation \citep{leitherer1999,lagarde2012}, and a
metallicity $Z = 0.002$ comparable to the one of the DLA ($\approx 0.1~\rm Z_\odot$).
We note that stellar population models predict an anti-correlation between metallicity and flux of ionizing radiation,
thus galaxy G will produce a lower $Q_{\rm H0}$ if it has a metallicity higher than what is observed in the DLA.
Assuming that a fraction $f_{\rm Ly\alpha} = 0.68$ of all 
recombinations gives rise to Ly$\alpha$ (in case B), galaxy G can account for a total luminosity of
$L_{\rm Ly\alpha,G} \approx 8.5\times 10^{42}~\rm erg~s^{-1}$, which is a factor of $\approx 3$ above the
total observed luminosity of the nebula, $L_{\rm Ly\alpha,neb} = (2.7\pm 0.1)\times 10^{42}~\rm erg~s^{-1}$.
However, if galaxy G emits isotropically and the nebular emission arises from $\approx 1/4$ of
the solid angle around the UV source, an escape fraction of $\approx 100\%$ in the direction of the nebula
is required to power Ly$\alpha$ with photoionization alone.

Our calculation is complicated by an unknown dust extinction, which is likely present inside this structure.
Indeed, the resonant Ly$\alpha$ emission
can be easily absorbed by dust, making the luminosity inferred for the nebula only a lower limit.
At the same time, galaxy G is likely to have an higher intrinsic star formation than what is inferred from
UV observations, making also our current estimate for $Q_{\rm H0}$ a lower limit.
As dust can easily absorb Ly$\alpha$ photons, as well as ionizing radiation, galaxy G appears to
be marginally sufficient for powering the observed Ly$\alpha$ emission.
Although potentially viable from an energy point of view, this interpretation requires a very high and possibly anisotropic
escape fraction, together with some degree of fine-tuning to reproduce the observed clumpiness inside the nebula.
It also does not naturally explain the apparent kinematic alignment between the absorption and emission, which would require
correlation of substructures within the CGM, but on opposite side of the galaxy star-forming region.
Thus, additional ionization mechanisms are likely responsible for, or contribute to, the observed Ly$\alpha$
emission.

\subsection{Contribution from {\em in situ} star formation}\label{sec:insitu}

A perhaps more natural explanation for the extended emission is the presence of {\em in situ} star formation
within the two clumps. It is in fact possible that clump \cE\ and \cW\ are actively forming stars
inside an extended gas rich structure, of which galaxy G is a member under our working assumption that
this galaxy lies at the DLA redshift. Indeed, after masking galaxy G,
the lack of extended UV continuum within the two clumps sets the rest-frame equivalent width to $>60.7$~\AA\
and  $>52.3$~\AA, respectively, implying that \cE\ and \cW\ are being powered by a recent burst of star
formation if the source of radiation is local (see below).
If \lya\ is powered {\em in situ}, star formation could be occurring inside a $\approx 40~\rm kpc$
proto-disk that is fragmenting in two clumps. However, the large separation of these clumps
would imply the existence of a very extended star-forming disk at $z\approx 3$, and therefore it appears more likely that star
formation occurs instead inside two low-mass galaxies that are on the verge of
merging \citep[e.g.][]{stierwalt2015,ribeiro2016}.  

This scenario is quantitatively consistent with the results of {\sc starburst99} calculations, in which
we follow the evolution of a single burst of $10^6~\rm M_\odot$ for $1~\rm Gyr$ using again the Geneva
stellar tracks with no rotation and with metallicity $Z = 0.002$ ($\approx 10\%$ solar). As before,
the Ly$\alpha$ luminosity is computed from the photon flux of ionizing photons
assuming case B recombination for gas at $T=10^4~\rm K$. 
The Ly$\alpha$ equivalent width is derived mimicking the observational techniques
by computing the UV continuum luminosity by averaging the stellar continuum over the wavelength
interval covered by the MUSE observations.
With this {\sc starburst99} model, we recover the well-known result of a decreasing
equivalent width as a function of time from the burst \citep[e.g.][]{charlot1993}, a trend
which we use to constrain the age of the burst.

By comparing our stellar population synthesis model to the inferred lower limits for the equivalent width
(which are independent on the assumed mass of the burst), we infer an upper limit on the age of the burst
of $\lesssim 7 ~\rm Myr$. At the same time, an estimate of the mass formed during this star formation event is
obtained by scaling the predicted Ly$\alpha$ luminosity (which is directly proportional to the mass of the burst)
to match the observed value. As the luminosity is a function of time, we conservatively
assume the maximum age of the burst allowed by the observed equivalent widths. With this comparison,
we find that the two bursts in \cE\ and \cW\ account for a total stellar mass of
$\approx 2\times 10^{7}~\rm M_\odot$ and $\approx 3\times 10^{7}~\rm M_\odot$, respectively
for the E and W clump. It should be noted that this simple mass estimate does not provide a
direct measurement for the total mass associated with the DLA in this structure, as it accounts only for stars
formed recently inside the two clumps under the assumptions described above.

A more extended star formation history
would indeed contribute to additional mass in stars, a possibility that can be tested with observations
in the rest-frame optical. Further, we emphasise that radiative transfer effects combined with an unknown
dust extinction make these estimates uncertain. However, our limits can be considered conservative, 
as plausible corrections would increase the equivalent width, resulting in an even younger and more massive burst.
Future deep observations in the far IR rest-frame will be able to assess the importance of dust-obscured
star formation in this system. 

Owing to more degrees of freedom, a model in which {\it in situ} star formation powers Ly$\alpha$ appears to
provide a more natural explanation for the observed clumps and for the kinematic alignment between emission and
absorption spectra. Indeed, the DLA absorption can be interpreted as extended gas in the CGM of the two
interacting galaxies, each of which retains the bulk of the halo systemic velocity.
The interaction between the two systems may also induce a tidal force that pushes neutral hydrogen to large distances
inside the CGM, as observed for instance in nearby interacting dwarfs \citep[e.g.][]{pearson2016}.

Gravitational interactions that remove enriched material from the interstellar medium of these galaxies can also account for
the observed metallicity of the DLA, which is significantly above the mean metallicity of the IGM and CGM at
these redshifts \citep[e.g.][]{schaye2003,mf2016lls}. Indeed, a $\approx 10\%$-solar metallicity resembles
the metal content of the interstellar medium of $\approx 10^{8.5}-10^{9}~\rm M_\odot$ galaxies according to the
$z \approx 3$ mass-metallicity relation \citep{mannucci2009}.
In addition to tidal interactions, galactic outflows are also
a plausible mechanism to enrich the DLA. However, a wind originating from the two bursts in clump \cE\ and \cW\
would have to travel at a velocity $v > 4000~\rm km~s^{-1}$ to reach the  DLA in  $< 7~\rm Myr$. Thus, it is more
plausible that older episodes of star formation, or winds from galaxy G and other galaxies below our detection limit,
have contributed to the observed enrichment.

Given the above discussions, it is clear that we cannot draw a firm conclusion on the nature of this
system based on current data. Our analysis, however, favours a model in which the quasar sightline is probing
an extended gas rich structure, within which two low-mass galaxies undergo a recent burst of star formation
triggered by their encounter. Galaxy G likely lies within the same structure, or at comparable redshift
within a group. Future observations targeting rest-frame optical lines and dust from these galaxies will
be able to further refine the relative position/velocities of the members of this structure, as well as
to refine our estimates on what powers the extended Ly$\alpha$ emission, thus providing more
definite information on this system. 

\section{Comparison with the \eagle\ simulations}\label{sec:cfreagle}

\begin{figure}
  \centering
  \includegraphics[scale=0.7]{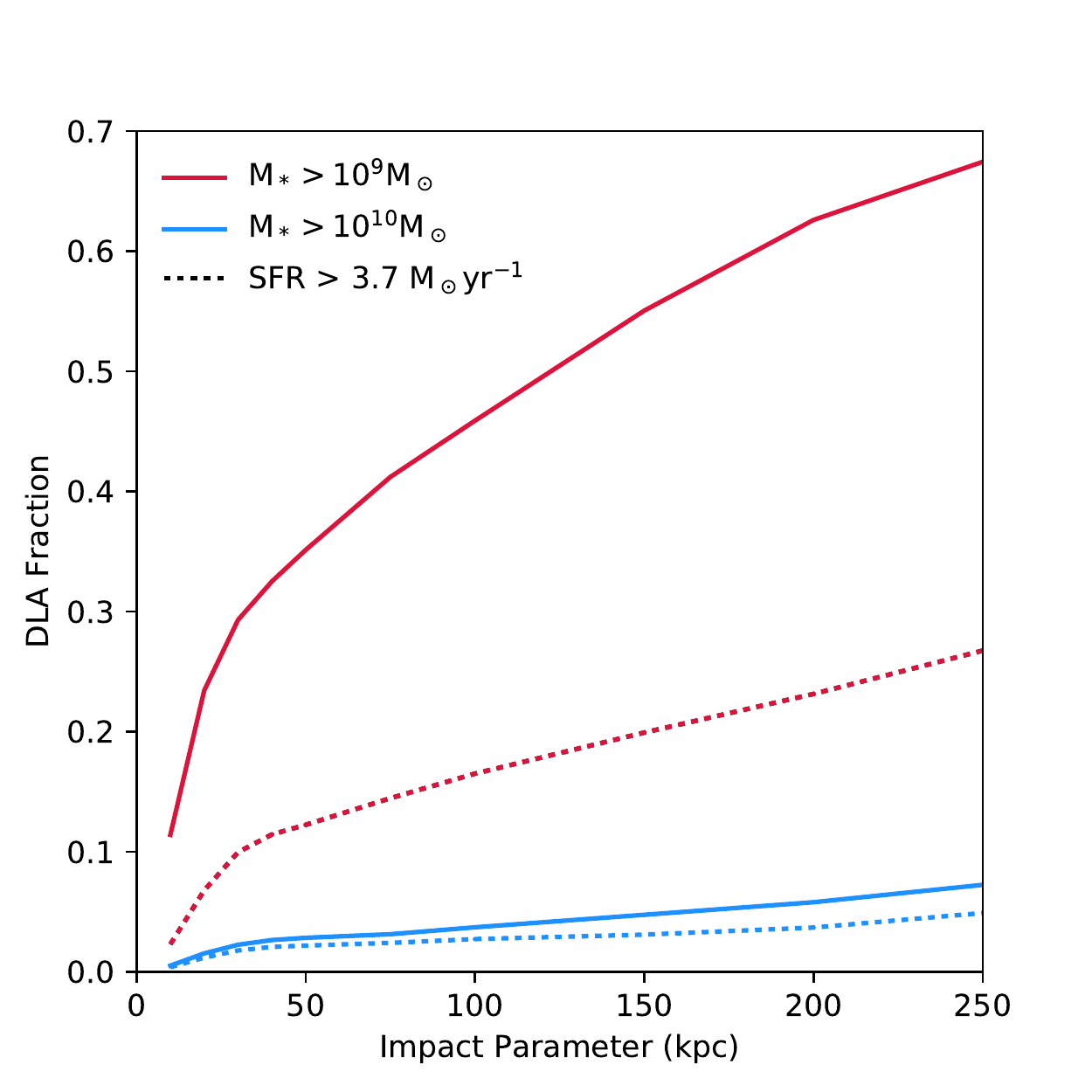}
 \caption{Fraction of DLAs in the \eagle\ L0025N0752 simulation with column density $\log (N_{\rm HI}/~\rm cm^{-2})\ge 20.55$ that intersect a galaxy at an impact parameter smaller than the value plotted along the horizontal axis, in proper kpc. Red and blue curves are computed requiring the galaxy to have a stellar mass greater than $10^9$ and $10^{10}$~M$_\odot$, respectively. The dashed lines are computed requiring that, in addition to the mass constraints, the SFR of the galaxy exceeds $3.7$~M$_\odot$~yr$^{-1}$.}\label{fig:EagleG}
\end{figure}

\begin{figure*}
  \centering
  \includegraphics[scale=0.55]{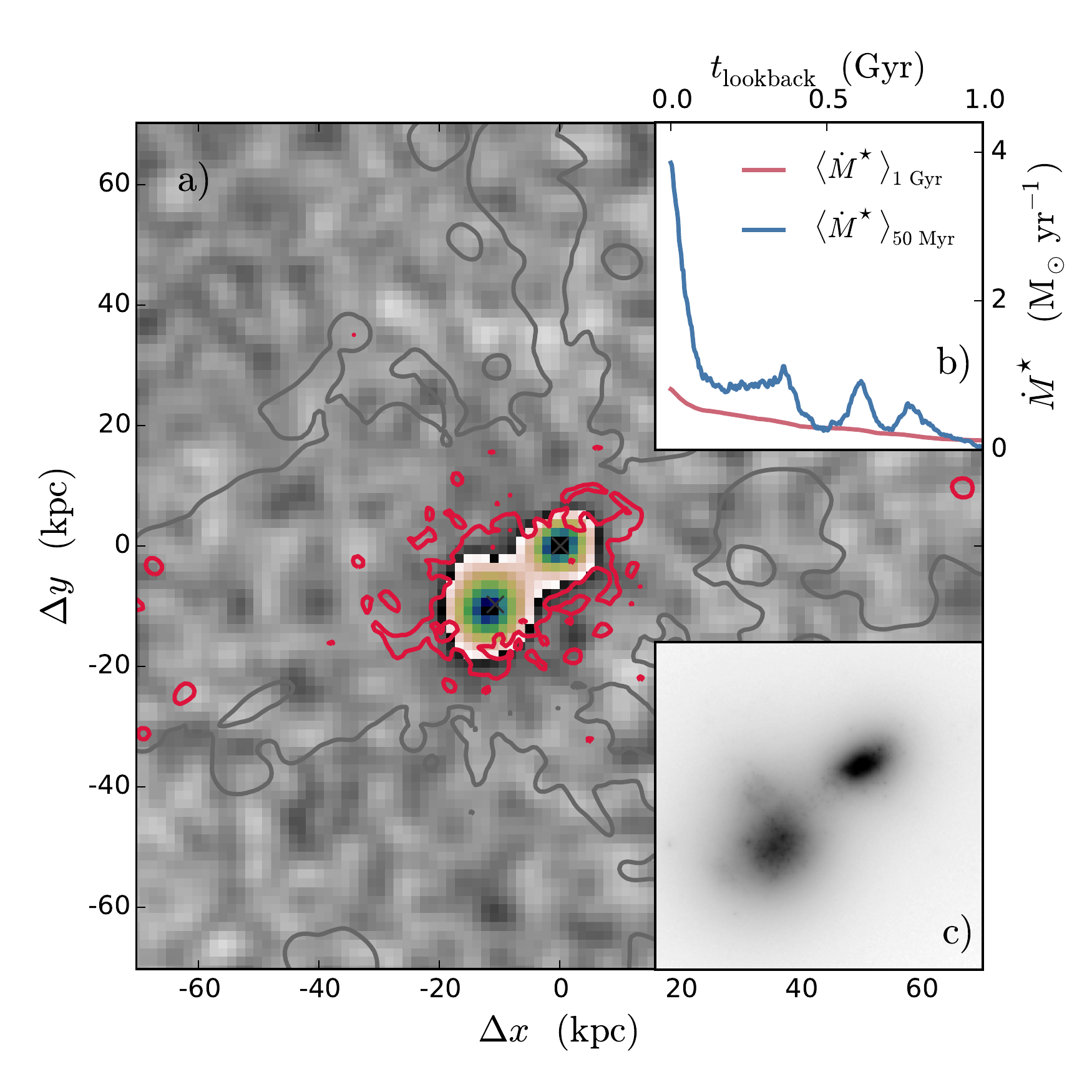}
  \caption{Mock observations of a system of two star-bursting galaxies associated with a DLA and exhibiting Ly$\alpha$ emission, extracted from the \eagle\ L0025N0752 simulation, with properties similar to the system observed in MUSE.
    {\it Panel a:} Mock Ly$\alpha$ image (with no radiative transfer),
    with the grey scale representing the noise layer matched to the MUSE data, revealing two bright
    Ly$\alpha$ clumps at a separation of $\approx 20~\rm kpc$. The colour map represents the
    intensity of the emission for an unresolved source.
    The red contour marks the {H\scriptsize~I} column density of $\log (N_{\rm HI}/~\rm cm^{-2})= 20.6$, showing that
    this system is embedded in an extended DLA with size comparable to that inferred by our observations.
    The grey contour at $\log  (N_{\rm HI}/~\rm cm^{-2}) = 16$ highlights how this simulated system is embedded in an extended gas filament. {\it Panel b:} The reconstructed star formation history of the two clumps combined, with $\dot M_\star$ averaged over the past 50~Myr and 1~Gyr shown in blue and red, respectively, illustrating that this
    system has a long-term star formation history consistent with the lack of UV detection in our observations, while it has
    experienced a recent burst that formed a stellar mass of $\approx 4 \times 10^{7}~\rm M_\odot$ (in agreement with observations) that could power the Ly$\alpha$ emission. {\it Panel c:} The $r$-band image reconstructed from the star formation history shows that  this system is undergoing a merger event, which induced the two synchronous bursts of star formation.}\label{fig:sims}
\end{figure*}

With only a handful of DLA sightlines being targeted by MUSE at present, we have limited empirical information on the frequency of extended nebulae near DLAs. To offer additional insight into the probability of uncovering similar systems in the future, we turn to the results of the \eagle\ cosmological simulations \citep{schaye2015,crain15}. A general prediction of these and other simulations is that DLAs are clustered with galaxies of a wide range of masses in $\Lambda$CDM \citep[e.g.][]{pontzen2008,rahmati2014}. Here, we focus on the more specific question of how common systems like the observed one are in simulated volumes at $z \sim 3$. To address this question, we start by asking what fraction of DLAs lies in proximity to galaxies similar to galaxy G. Next, we estimate how common DLAs are in group environments, concluding with a search of a close-analogue to the observed system.

\subsection{The \eagle\ simulations and radiative transfer post-processing}

\eagle\ is a suite of cosmological hydrodynamic simulations performed in cubic volumes with co-moving lengths ranging from $L=12.5$ to 100 co-moving Mpc, with cosmological parameters taken from the 
\cite{planck14} analysis. Simulations are performed using the {\sc Gadget-3} incarnation of the tree-SPH code described by \cite{springel05}, with modifications to the hydrodynamics and time-stepping schemes (see \cite{schaller15} for details), and the inclusion of a range of subgrid models to account for unresolved physics below the resolution scale, calibrated as described by \cite{crain15}. Briefly, the simulations incorporate 
the major ingredients required to model galaxy evolution, 
including element-by-element radiative cooling and photoheating, a pressure-dependent star formation law above a metallicity-dependent density threshold, mass loss from stars and supernovae, and thermal feedback associated with star formation and active galactic nuclei\footnote{Details on these implementations can be found in \cite{schaye2015}, together with relevant discussion on differences with previous simulations, e.g. OWLS \citep{schaye2010}.}.
	
Here we use two different \eagle\ runs from table~2 in \cite{schaye2015}. To examine the correlation between galaxies and DLAs
	we use the higher resolution run L0025N0752 ($L=25$~Mpc, SPH particle mass $m_g=2.26\times 10^5$~M$_\odot$, co-moving Plummer-equivalent softening $\epsilon_{\rm com}= 1.33$~kpc).  To investigate the presence of multiple galaxies in the same structure, we use instead run L0100N1504 ($L=100$~Mpc, $m_g=1.81\times 10^6$~M$_\odot$, $\epsilon_{\rm com}= 2.66$~kpc), which has better galaxy statistics. Relevant to the current analysis is that \eagle\ reproduces the $z\approx 3$ galaxy stellar mass function to within current observational uncertainties, although the specific SFRs are low by $0.2-0.5$~dex if the data are taken at face value \citep{furlong15}. 

We use the {\sc urchin} reverse-ray tracing code described by \cite{altay2013} to compute the neutral fraction of all gas particles, assuming the IGM to be photoionized at the rate computed by \cite{haardt01}. 
Projecting the neutral hydrogen density of all gas particles onto a side of the simulation box allows us to evaluate the \ion{H}{I} column density for $8192^2$ sightlines parallel to a coordinate axis of the cube. Applied to \eagle,
	this procedure reproduces the observed number of DLAs (as well as Lyman limit systems) per unit absorption distance as a
	function of column density very well \citep[see][]{rahmati2015}, at least as well as the {\sc owls} simulation analysed in the same way by
	\cite{altay11}. Galaxies in \eagle\ are identified as self-bound substructure inside dark matter halos, using the {\sc subfind}
	algorithm of \cite{springel01} and \cite{dolag09}. Next, we compute the stellar mass and SFR for each galaxy, the
	position and peculiar velocity of their centres of mass, as well as the mass of their parent dark matter halos \citep[see][]{mcalpine2016}. Finally, we correlate DLAs to galaxies, by combining the \ion{H}
	{I} column through the \eagle\ volume with the location of galaxies. 
	
We note how, in the previous section, we have explored 
the relevance of ionizing radiation produced by local sources in powering the observed Ly$\alpha$ emission in the nebula. However, when comparing with the results of the \eagle\ simulations, we do not include photoionization from local sources for the following reason. While local radiation is clearly relevant to model in detail the particle-by-particle ionization state of the gas and its emissivity near galaxies, in practice, the inclusion of radiative transfer effects from local sources is subject to uncertainties related to the unresolved structure of the interstellar medium within large cosmological simulations. These uncertainties are particularly significant at high column densities \citep[$N_{\rm HI} \gtrsim 10^{21}~\rm cm^{-2}$;][]{rahmati2013}, which are relevant for this work. 

Therefore, we refrain from deriving predictions of the \lya\ luminosity from DLAs in \eagle, focusing instead on the more basic task of comparing the general DLA population with galaxies in the \eagle\ volume. As discussed in the literature \citep[e.g.][]{mf2011,rahmati2013}, while local sources of radiation may alter the  cross section of individual absorbers, extended DLAs next to galaxies persist even after including local sources. This implies that associations between DLAs and galaxies in simulations are generally robust even without inclusion of local sources during the radiative transfer post-processing. Nevertheless, in the following, we  provide a qualitative description of how sensitive our conclusions are to the assumptions made during radiative transfer post-processing.

\subsection{Statistical analysis of DLAs associated to galaxies in \eagle}

We estimate the probability with which systems like the one uncovered by our study can be found in future observations by first quantifying the fraction of DLAs that resides near galaxies with properties similar to galaxy G (with the caveat on the redshift discussed above). For this, we begin by identifying all DLA sightlines with $\log (N_{\rm HI}/~\rm cm^{-2})\ge 20.55$, which is the minimum column density of the observed DLA as discussed in Sect.~\ref{sect:DLA}. 
The likelihood of observing a galaxy with properties similar to that of galaxy G (see Sect.~\ref{sect:galaxies}) within a given impact parameter from a DLA with this column density is shown in Figure~\ref{fig:EagleG}.

The observed projected distance between the DLA and galaxy G is $\approx 20$~kpc, but the galaxy stellar mass is relatively uncertain, and therefore we show curves for two choices, $M_\star=10^9$ and $10^{10}$~M$_\odot$ (red and blue solid
	curves, respectively). From this analysis, it is clear that finding a DLA-galaxy pair with such a close impact parameter has a probability of  $\approx 23\%$ and $\approx 2\%$, respectively for the two choices of mass.
        We note that this probability would further increase if we had to consider a higher column density for the DLA.
        When we require that the galaxy undergoes star formation at a rate exceeding $3.7~\rm M_\odot~yr^{-1}$, we find that
        these probabilities drop to $7\%$ and $1\%$, respectively  (dashed lines in Figure~\ref{fig:EagleG}). 

As discussed above, since we do not include local sources in our calculation, the total number of DLAs recovered in this calculation is formally an upper limit. However, DLAs sightlines are clustered in structures with a range of column densities, implying that neutral gas with column densities above the $\log (N_{\rm HI}/~\rm cm^{-2})\ge 20.55$ threshold is likely to be found in the same regions identified by our study even when radiative transfer effects from local sources are included. For this reason, we regard our conclusions as largely insensitive to the details of the radiative transfer post-processing. 

Next, we investigate how frequently a DLA sightline with $\log (N_{\rm HI}/~\rm cm^{-2})\ge 20.55$ passes near a galaxy with properties similar to those of clumps \cW\ and \cE\ combined. In this case, we require that the SFR averaged over the past $10~\rm Myr$ is $\dot M_\star(10~{\rm Myr})>3$~M$_\odot$~yr$^{-1}$, whereas when averaged over the past 1~Gyr is $\dot M_\star(1~{\rm Gyr})<5$~M$_\odot$~yr$^{-1}$. This is in line with our estimates from Sect.~\ref{sect:galaxies}. For the observed impact parameter between the clumps and the DLA of $30$~kpc, we find that these criteria apply to $\approx 3\%$ of the selected DLA sightlines in \eagle. Again, we regard these statistics as largely independent of radiative transfer effects for the reasons discussed above.
	
Finally, we use \eagle\ to quantify the joint probability of finding a galaxy with properties consistent with those of galaxy G near two star-forming galaxies consistent with \cW\ and \cE. For this calculation, we turn to the $L=100~\rm Mpc$ box to increase the statistical significance of this analysis. To mimic our observations, we require that one galaxy has $M_\star\ge 10^9~\rm M_\odot$ (as galaxy G), which is within a projected separation $\le 50$~kpc of two star-bursting galaxies with $\dot M_\star(10~{\rm Myr})>2$~M$_\odot$~yr$^{-1}$, in line with the inferred properties of the two clumps which form $\approx 2\times 10^{7}~\rm M_\odot$ of stars in $\approx 10~\rm Myr$.  
To match the observed equivalent widths, we further require that $\dot M_\star(1~{\rm Gyr})<5$~M$_\odot$~yr$^{-1}$.
Finally, we demand that the two galaxies are less than $30~\rm kpc$ apart (projected distance), to mimic
the observed separation between \cE\ and \cW.

After imposing these constraints, we find 21 such triples in the \eagle\ run with $L=100~\rm Mpc$. In most of the cases, the triple corresponds to a situation where all three galaxies are members of the same galaxy group, with halo mass in the range $10^{12}-10^{13}$~M$_\odot$. We note that for each pair (\cE\ and \cW),
there are several nearby galaxies that are consistent with galaxy G, implying that other star-forming galaxies
are predicted in the surroundings of DLAs associated with these groups.
 The velocity offset between these galaxies ranges between $100-500~\rm km~s^{-1}$, consistent with the virial velocity\footnote{We employ the usual definition where $R_{200}$ is the radius of a halo, defined such that the mass within $R_{200}$ equals $M_{200}=200\times (4\pi/3)\rho_c\,R_{200}^3$, where $\rho_c$ is the critical density.} of
\begin{equation}
  V_c=(10{\rm G}H(z)M_{200})^{1/3}=240~{\rm km}~{\rm s}^{-1}\,(M_{200}/10^{12}{\rm M}_\odot)^{1/3},
\end{equation}
  in groups of this halo mass at this redshift. Such offsets are consistent with those found between galaxy G, the clumps, and the DLA. We conclude from this that the observed system is consistent with the DLA being associated with a galaxy group, of which G, \cE\ and \cW\ are members under our working hypothesis that $z_{\rm G} = z_{\rm dla}$. Such groups are relatively rare at $z=3.25$, with a number density of $\approx 10^{-4}$ per comoving Mpc$^{3}$, with typical radius of $R_{200}=100$~physical kpc, a size which is consistent with the observed projected separations. Only a small fraction of such groups contain two star bursting and merging galaxies of the types of the observed clumps.

Altogether, we conclude from this comparison with the results of the \eagle\ simulation that systems such as the one observed - consisting of a DLA, with a nearby relatively massive galaxy and a further two lower mass star-bursting galaxies - do occur in simulations, although not with high number density.  In the majority of cases, such systems are associated with a group of galaxies, with the DLA also associated with the group. The velocity offsets amongst the galaxies themselves, and between the galaxies and the DLA, are predicted to arise from virial motions in a dark matter halo of mass $10^{12}-10^{13}~{\rm M}_\odot$.
This result implies that, for the case of groups, individual galaxy-absorber associations cannot be trivially established, and that these do not provide a complete description of the systems probed in absorption.

\subsection{Close analogue of the observed system}

We conclude this section by briefly discussing the properties of a close analogue of the observed association of two star bursting galaxies near a DLA, which we have identified during this analysis within the \eagle\ L0025N0752 run. This simulated object offers an example that configurations like the observed one - with two galaxies in synchronous star-burst phases inside an extended DLA - are plausible within the current cosmological model.

The simulated system, which is shown in Figure~\ref{fig:sims}, is composed of two galaxies that are undergoing a merger that triggers two synchronous bursts of star formation, and which form a total of $\approx 4\times 10^{7}~\rm M_\odot$ of stars within $\approx 10~\rm Myr$, similar to the observed clumps \cW\ and \cE. Prior to this encounter, the two galaxies were forming stars at a rate $\lesssim 1~\rm M_\odot~yr^{-1}$, and hence, again similar to clumps \cW\ and \cE, would not have been detected in our current
	UV continuum observations. Inset {\it b} shows the recent star formation history of both clumps combined, measured back in time with look back time $t=0$ chosen to correspond to the time at which the system is observed in the simulation. This close analogue is further embedded within a $\log (N_{\rm HI}/~\rm cm^{-2}) \ge 20.6$ DLA with an extent of $\gtrsim 40~\rm kpc$ (red contours), which is part of a large-scale filament, as traced by lower column density neutral hydrogen (grey contours). Being related to the properties of a single object, we caution that this part of the analysis is more 
prone to uncertainties introduces by our assumptions on 
radiative transfer. The simulation further predicts a mass-weighted metallicity for this system of about $\approx 10\%$ solar, again close to the observed value.

Using the star formation history known from the simulation, we generate a mock Ly$\alpha$ image (background grey scale in the main panel) with depth and resolution comparable to the MUSE data, by scaling the Ly$\alpha$ luminosity from the stellar population synthesis model described above according to the stellar mass generated within the simulation during the burst. As before, we neglect radiative transfer effects and dust absorption. The resulting map is computed for the Ly$\alpha$ luminosity close to the peak of the burst (at $< 1~\rm Myr$). We also generate a rest-frame $r$-band image of the stellar continuum at full resolution following the method discussed in \citet{trayford17}, shown in panel {\it c}.

\section{Summary and Conclusions}\label{sec:disc}

We have presented new MUSE observations of the quasar field J$025518.58+004847.6$  ($z_{\rm qso}\approx 3.996$),
which is known to host an intervening DLA at $z_{\rm dla} = 3.2552 \pm 0.0001$ with
$\log N_{\rm HI} > 20.55 \pm 0.10 ~\rm cm^{-2}$ (with a best estimate of $\log N_{\rm HI} = 20.85 \pm 0.10 ~\rm cm^{-2}$)
and metallicity $\log Z/Z_\odot < -0.8 \pm 0.1$ (with a best estimate of $\log Z/Z_\odot = -1.1 \pm 0.1$).
By reaching a flux limit of $> 2\times 10^{-18}~\rm erg~s^{-1}~cm^{-2}$ ($3\sigma$), our observations
uncover a $37\pm 1~\rm kpc$ extended structure that emits in Ly$\alpha$ with a total luminosity
of $(27 \pm 1)\times 10^{41}~\rm erg~s^{-1}$. The centre of this nebula is located at a projected distance of
$30.5 \pm 0.5 ~\rm kpc$ from the quasar sightline.

This nebula also contains two distinct clumps, for which no continuum is detected to rest-frame equivalent
limits of $>60.7$~\AA\ and  $>52.3$~\AA. The two clumps are separated by $16.5\pm 0.5~\rm kpc$
in projection and have a line-of-sight velocity difference of $\Delta v_{\rm E-W} = 140\pm 20~\rm km~s^{-1}$,
which is consistent with the velocity difference of $\Delta v_{1-2} = 155 \pm 6~\rm km~s^{-1}$ measured
in between the two main absorption components seen in the metal transitions associated with the DLA.

Furthermore, a compact galaxy is detected in the continuum at a projected distance of $19.1 \pm 0.5 ~\rm kpc$
from the quasar sightline. The MUSE spectrum of this source is noisy, but with properties consistent with a $z \approx 3.2$
star-forming galaxy. Given its close proximity to the nebula and its photometric properties, we consider this galaxy to be physically associated with the structure hosting the DLA. 

Based on current observations, two mechanisms appear plausible to power
the extended Ly$\alpha$ emission. First, with a star formation rate of $\approx 4~\rm M_\odot~yr^{-1}$,
photoionization from the continuum-detected galaxy may ionize the galaxy CGM, powering the nebula.
However, this scenario requires an elevated escape fraction ($\approx 100\%$) of ionizing radiation from the sites
of star formation, although an unknown dust extinction hampers robust conclusions. 
A second viable model for powering the nebula is that Ly$\alpha$ emission arises from {\em in-situ} star formation
inside the two embedded clumps, in the form of two young ($\lesssim 7~\rm Myr$) starbursts likely triggered by a
merger of two galaxies. This scenario also provides a satisfactory explanation, at least qualitatively,
for the observed kinematic alignment between the two clumps and the  main absorption components of the DLA.
Finally, a combination of {\em in-situ} star formation and ionization from galaxy G may also account
for the total energy budget. Additional follow-up observations at rest-frame optical and IR wavelengths are now required to
better characterise the nature of this system, especially by confirming the redshift of the
continuum detected galaxy and by constraining the presence of dust inside this structure.

Regardless of what is powering the observed emission, the position of the DLA and the extent of the
Ly$\alpha$ nebula jointly suggest the presence of a gas-rich structure that stretches for
$\gg 50~\rm kpc$ on a side, inside which one or multiple galaxies are forming stars. This structure is
reminiscent of the filaments predicted by hydrodynamic cosmological simulations, at the intersection of
which galaxies form. Within this model, we would expect a fraction of DLAs to be embedded
within structures hosting galaxies that are surrounded by extended halos or filaments
which radiate in Ly$\alpha$ emission \citep{cantalupo2012,rauch2013,wisotzki2016}.

The association between extended Ly$\alpha$ nebulae discovered in deep long-slit spectroscopy or narrow-band
images and the population of optically-thick absorbers, and particularly DLAs, has already been explored
from a statistical point of view by \citet{rauch2008}.
This study suggests that low-surface brightness nebulae have sufficiently large number density and
covering factor to account for the observed incidence of DLAs.
Compared to the properties of the nebulae presented in this work, however, the emitters uncovered in the
deep long-slit observations by \citet{rauch2008} are generally more compact, with sizes
$\lesssim 20~\rm kpc$ measured at $10^{-19}~$\sbline. Only a few cases ($4/27$ objects)
are found with sizes $\gtrsim 30~\rm kpc$, which is similar to the extent of the 
nebula discovered in our MUSE observations. Moreover, the objects from \citet{rauch2008} have a
typical luminosity $\lesssim 10^{42}~\rm erg~s^{-1}$, again placing the nebula near the $z\approx 3.25$
DLA in the extreme tail of the population of these emitters.

The Ly$\alpha$ sources discussed in \citet{rauch2008} represent ``field'' objects which are not explicitly associated to
a DLA. However, examples of diffuse Ly$\alpha$ nebulae in proximity of DLAs, which exhibit characteristics similar
to the system described in this work, can be found in the literature \citep[e.g.][]{fynbo1999,zafar2011,kashikawa2014}.
In particular, while the system reported in \citet{fynbo1999} and \citet{zafar2011} lies in proximity to 
a quasar, the object presented by \citet{kashikawa2014} is a close analogue to the system
discussed here, being associated to an intervening DLA with a luminosity of $\approx 10^{42}~\rm erg~s^{-1}$
and a size of $\approx 15~\rm kpc$ as measured in a 2D spectrum at a depth of
$2.5\times 10^{-19}~\rm erg~s^{-1}~cm^{-2}~\AA^{-1}$.

A question that arises from our study and previous work
is how common extended nebulae near DLAs are. Our discovery of a $\approx 40~\rm kpc$
nebula in a field that had been studied previously with optical imaging campaigns yielding no detections
\citep{mf2014,mf2015} hints that previous imaging and, to some extent, spectroscopic observations have been 
unable to identify these types of associations with high completeness.
Hence, our study offers a clear example of the need for deep
multiwavelength campaigns to map simultaneously the galaxy population and the more diffuse gas environment near DLAs.

Based on the number density estimates in \citet{rauch2008} and our comparisons with the \eagle\ simulations, it is however unlikely that nebulae with sizes
$\gtrsim 30~\rm kpc$ and luminosity $\ge 10^{42}~\rm erg~s^{-1}$ are common occurrence next to DLAs.
Nevertheless,  this system offers a tantalising example of how continuum-detected sources \citep[especially
  near metal-rich DLAs;][]{krogager2017} may not represent isolated associations but trace instead
clustered galaxy formation inside a web of filaments. In fact, the presence of DLAs in interacting groups has already been
confirmed with observations in the literature. For instance, \citet{warren1996} and \citet{weatherley2005} reported
the detection of multiple emission-line galaxies in the fields of two DLAs at $z\approx 2$ and $z\approx 3$.
The plausibility of DLAs being associated with clustered galaxy formation is also confirmed by our analysis of the
\eagle\ simulations.

At present, we cannot derive firm conclusions on the number density of these types of associations,
as a systematic exploration of the incidence of extended nebulae and groups in proximity to DLAs
has not been possible so far due to the difficulties of reaching flux limits of
$10^{-18}~\rm erg~s^{-1}~cm^{-2}$ \citep{christensen2007}. However, the deployment of sensitive large-format integral
field spectrographs at 8-meter telescopes and of the full ALMA
array will soon enable deep and complete multiwavelength searches for both
star-forming galaxies (including dust-obscured systems) and diffuse emission near DLAs. These observations, in tandem with
refined model predictions of the emission properties from dense filaments that are traced by strong absorption line systems,
will offer a novel way to test if galaxy formation proceeds inside gas-rich filaments as predicted by current cosmological
simulations.

\section*{Acknowledgements}

We thank G. Ashworth for producing preliminary stellar population synthesis models and R. Cooke for assistance with the modelling of absorption lines. We also thank J. Schaye for providing comments on a draft of this manuscript, and the \eagle\ team for making the comparison
with simulations possible. M.F. and T.T. acknowledge support by the Science and Technology Facilities Council [grant number ST/P000541/1]. T.T. also acknowledges the Interuniversity Attraction Poles Programme initiated by the Belgian Science Policy Office ([AP P7/08 CHARM]). JXP acknowledges support from the National Science Foundation (NSF) grants AST-1010004 and AST-1412981. This work is based on observations collected at the European Organisation for Astronomical Research in the Southern Hemisphere under ESO programme ID 095.A-0051. Some of data presented herein were obtained at the W.M. Keck Observatory. Part of this work used the DiRAC Data Centric system at Durham University, operated by the Institute for Computational Cosmology on behalf of the STFC DiRAC HPC Facility (www.dirac.ac.uk). This equipment was funded by BIS National E-Infrastructure capital grant ST/K00042X/1, STFC capital grant ST/H008519/1, and STFC DiRAC is part of the National E-Infrastructure. This research made use of Astropy, a community-developed core Python package for Astronomy \citep{astropy}. For access to the data and codes used in this work, please contact the authors or visit \url{http://www.michelefumagalli.com/codes.html}. VLT data are also available via the ESO archive.
 









\bsp 
\label{lastpage}
\end{document}